\documentclass[final]{aipproc}

\layoutstyle{8x11double}

\begin{document}

\title{Updated axion CDM energy density}


\author{Ji-Haeng Huh}{email={jhhuh@phya.snu.ac.kr}
, address={Department of Physics and Astronomy and Center
for Theoretical Physics, Seoul National University, Seoul 151-747, Korea}}

\begin{abstract}
We update cosmological bound on axion model. The contribution from the anharmonic effect and the newly introduced initial overshoot correction are considered. We present an explicit formula for the axion relic density in terms of the QCD scale $\Lambda_{QCD}$, the current quark masses $m_q$'s and the Peccei-Quinn scale $F_a$, including firstly introduced $1.85$ factor which is from the initial overshoot.
\end{abstract}

\maketitle

\section{Introduction}
The standard model suffers several fine tuning problems. They are all related to the symmetry of the nature and its breaking. The gauge hierarchy problem which is one of the main motivation of the upcoming experiment, LHC has its origin in the gauge symmetry breaking sector. Another problem, strong CP problem is deeply related to a discrete symmetry, CP. The fact that the almost of the suggestions to solve them are more or less related to the invisible missing matter in the Universe is interesting.

The strong CP problem can be stated as that a CP-odd operator $\bar\theta tr\{G\tilde G\}$ which cannot be forbidden is unmeasurably small $\sim10^{-11}$ \cite{Baluni79}. The introduction of the axion can solve this problem beautifully \cite{PQ77,Kim79}. The axionic solution introduces an irrelevant operator $\frac{a}{F_a}tr\{G\tilde G\}$. It can be thought of as making $\bar\theta$ a dynamical field. If then, the Vafa-Witten theorem tells that CP conserving point $a+\bar\theta=0$ becomes a minimum of the free energy by the strong interaction \cite{Vafa:1984xg}.

However, the introduction of the axion coupling solely cannot solve the strong CP problem. One needs one more ingredient, the relaxation mechanism for axion. The standard big bang cosmology provides it naturally, so the axionic solution to the strong CP problem is intrinsically cosmological one. This point was recognized soon after the suggestion of the axion \cite{Preskill83}. The axion is one of the possible candidate of the cold dark matter.

After the early studies of cosmological implication of the axion, Turner \cite{Turner86} made a quantitative, numerical estimate of the axion energy density with a anharmonic correction and more precise axion mass formula.

All of their work depends on the assumption of the adiabatic process. However, the assumption can be broken in the two phases of the evolution, at the time of starting to roll down and at the time of QCD phase transition \cite{DeGrand:1984uq,DeGrand:1985uq}.

We considered the effects of these two phases on the axion energy density and reanalyzed anharmonic effect. In addition, many parameters of QCD physics and the cosmology are changed \cite{Manohar06,Hinchliffe02}. We finally represent the explicit formula in terms of these parameters.

\section{relic density of axion}
Just after the inflation, a very light scalar field like axion tends to be frozen at a certain value \cite{Linde82}, $\langle \Phi\rangle=\phi_0={\rm const.}$, which is determined stochastically. The dynamics of a scalar field in the expanding universe with the Hubble parameter $H(t)$ is governed by the equation of motion $\ddot\Phi+3H(t)\dot\Phi+V'(\Phi)=0$. The second term in the left hand side of it prevents roll down of the scalar field, while second term forces it to move to the minimum. Therefore, after the inflation, the scalar fields stays at their initial value $\phi_0$ until the forces $V''$ to roll down overcome the Hubble friction $3H\dot\Phi$.

If the scalar potential is harmonic one, it starts rolling down and oscillates when $3H$ becomes small as the mass $m$. In the hydrodynamic description, such a coherently oscillating scalar fields behaves as like the pressureless cold($T=0$) matter. If the interaction with the known particles is sufficiently weak and its mass is light enough, it can be a cold dark matter(CDM) candidate. Since the invisible axion has the interactions and the mass originating from quantum effect, they can be small enough. So it's a natural candidate of this kind CDM.

The potential of the axion field can be parameterized by $V(\theta)=-C(T)\cos(\theta)$, where
$\theta=a/F_a$ and $C(T)=\alpha_{\rm inst}{\rm GeV}^4 (T/1{\rm GeV})^{-n}$. If the initial miss alignment angle $\theta_1$ is small enough, it becomes the harmonic potential with the mass $m_a(T)=\sqrt{C(T)/F_a^2}$. At the temperature $T_1$ satisfying $m_a(T_1)\simeq 3H(T_1)=1.66g_*^{1/2}T^2/M_P$, the axion field starts rolling down. In our parametrization, $T_1=\left(\frac{\alpha_{inst}}{4.98 g_*^{1/2}}\frac{M_P}{F_a}\right)^{2/(4+n)}{\rm GeV}$. Under the adiabatic assumption, $H,\dot m_a/m_a \ll m_a$, the number of the axion is conserved. So the axion energy density at present can be easily calculated as $\rho_a\simeq m_a(T_{\gamma})m_a(T_1)\biggl(\frac{R^3(T_1)
}{R^3(T_{\gamma})}\biggr)\theta_1^2$.

\section{temperature dependent axion mass}
To give a prediction, we should know $\alpha_{\rm inst}$ and $n$. Although, we cannot solve it fully, all we have to know is the mass in the two different regime in which we can estimate it numerically or phenomenologically. Two regimes are zero temperature hadronic phase and the temperature enough higher than $\Lambda_{\rm QCD}$ \cite{Kim:2008hd}.

At the zero temperature axion mass is given by $\sqrt{Z/(1+Z)}m_\pi f_\pi/F_a$, where $Z=m_u/m_d$. At the high enough temperature, axion mass can be obtained by the perturbative  instanton calculus under the dilute gas approximation \cite{Yaffe81}. The instanton density with the size $\rho$ is
\begin{eqnarray}
n(\rho,T)&=&m_um_dm_s C_N(\xi\rho)^3\frac{1}{\rho^5}
\left(\frac{4\pi^2}{g^2}\right)^{2N}
e^{-8\pi^2/g^2(\Lambda)}\nonumber \\
&&\times e^{\left\{-\frac13\lambda^2 (2N+N_f) -12 A(\lambda)\left[1+\frac16(N-N_f)\right]\right\}},\nonumber
\end{eqnarray}
where $\xi= 1.33876$ and $C_N=0.097163$ for $N=3$. With the 3-loop running coupling constant, the numerical results are well fitted at the $T\sim1{\rm GeV}$ by the form \cite{Bae:2008ue}
\begin{eqnarray}
\alpha_{\rm inst}=&3.96\times 10^{-12} \left(\frac{m_um_dm_s}{3\cdot 6\cdot 103~\rm MeV^3}
\right)\left(\frac{\Lambda_{\rm QCD}}{\rm 380~ MeV}\right)^{7.97},\nonumber\\
n=&7.44259-0.564492 \left(\frac{\Lambda_{\rm QCD}}{\rm 380~ MeV}\right).\nonumber
\end{eqnarray}

\section{possible corrections}
Our rough estimation relies on two assumptions, the harmonic limit($\theta_1\ll1$) and the adiabatic approximation($\dot m_a/m_a,H\ll m_a$). We considered possible effects of them on the axion energy density.

There is also a possible correction from the QCD phase transition. However, it turns out to be negligible, if one does not include fine tuning \cite{Bae:2008ue,DeGrand:1985uq}.

\subsection{anharmonic correction}
In the previous section, we use the fact that the axion number is conserved. However, the axion field is in the coherent state which is not the number eigenstate. Therefore the concept of the number conservation is not obvious except for the case of the non-interacting harmonic limit. The conservation quantity, total number of axion is actually an adiabatic invariant of the system under the slowly varying potential. In the anharmonic limit, $\theta\sim 1$, there is no interpretation like the number conservation.

The adiabatic invariant is given by the area in the phase space swept by one period, $\oint p~dq$ \cite{Landau}. In our case, its effect is appeared as a correction factor \cite{Bae:2008ue,Lyth91} $f_1(\theta_1)=\frac{2\sqrt{2}}{\pi\theta_1^2} \int^{+\theta_1}_{-\theta_1}\sqrt{\cos\theta'-
\cos\theta_1} \; d\theta'$. Now the axion energy density with the anharmonic correction is $\rho_a\simeq m_a(T_{\gamma})m_a(T_1)\biggl(\frac{R^3(T_1)
}{R^3(T_{\gamma})}\biggr)\theta_1^2 f_1(\theta_1)$.

\subsection{initial overshoot correction}
There are two adiabatic conditions we assumed. One is for the expansion rate to be much smaller than the axion oscillation, $H\ll m_a$. Another is for the change rate of the mass to be smaller than the oscillation. However, at the time of starting to roll down, the previous one cannot be satisfied because the expansion rate is comparable with the mass, $3H\simeq m_a$ at that time.

Therefore the adiabatic approximation does not work in the first few oscillations, and the numerical analysis must be done in this period. To include its effect in our calculations, we defined the another temperature $T_2$ and the new misalignment angle $\theta_2$ at the temperature $T_2$, such that the adiabatic condition is well satisfied after $T_2$. We explicitly checked that the values after the first half oscillation are good enough. We obtained $T_2/T_1=f_2(\theta_1,n)$ and $\theta_2\theta_2=f_3(\theta_1,n)$ numerically, and they are presented in the fig.

As one can sees, the correction factors, $f_2$ and $f_3$, becomes larger at the large $\theta_1$. It is because of that the anharmonic terms make the field stays longer time at the around of the top of the potential. So this initial overshoot correction includes other anharmonic correction. To split them, it is useful to factor out overall constant as like $f_2(\theta_1,n)\simeq c_2 \tilde f_2(\theta_1,n)$ and $f_3(\theta_1,n)\simeq c_2 \tilde f_3(\theta_1,n)$, where the function with a tilde goes to $1$ as $\theta_1$ is going to zero.

\begin{figure}[ht]
\centering
\resizebox{0.85\columnwidth}{!}
{\includegraphics{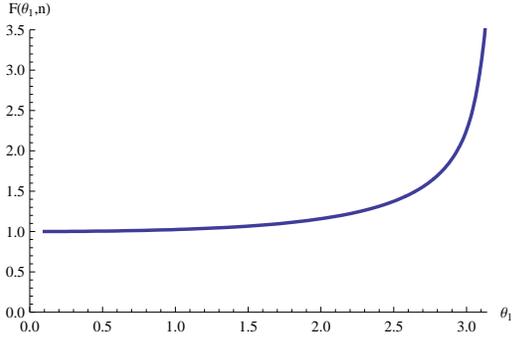}}
\caption{The combined correction factor $F(\theta_1,n)$.}\label{fig:combincorr}
\end{figure}

With the corrections $f_1$, $f_2$ and $f_3$, the relic density becomes
\begin{eqnarray}
\Omega_a &\propto& \frac{m_a(T_2)\theta_2^2}{T_2^3}f_1(\theta_2)\nonumber\\
&\propto&\frac{\theta_1^2c_3^2\tilde f_3^2(\theta_1,n)f_1(\theta_1c_3\tilde f_3(\theta_1,n))}{c_2^{3+n/2}\tilde f_2^{3+n/2}(\theta_1,n)}\nonumber\\
&=& 1.846\times\theta_1^2\times F(\theta_1,n).\nonumber
\end{eqnarray}
where the function
\begin{equation}
F(\theta_1,n)=\tilde{f}^2_3(\theta_1,n)f_1(\theta_1c_3\tilde{f}_3(\theta_1,n))/\tilde{f}_2^{3+n/2}(\theta_1,n) \end{equation}
which is shown in Fig. \ref{fig:combincorr}. As one can see, there is a $1.846$ enhancement in the axion relic density.

\section{results and conclusion}

With all of corrections we considered, the relic density of the axion is estimated as
\begin{equation}
\rho_a\simeq 1.449 \times 10^{-11} \biggl(\frac{F_{a, \rm GeV}}{10^{12}} \biggr) \frac{1}{T_{1, \rm GeV}}\frac{\theta_1^2}{\gamma} F(\theta_1,n) (\rm eV)^4,\nonumber
\label{axion:density}
\end{equation}
where we used $Z\equiv m_u/m_d = 0.5$, $m_{\pi^0}=$135.5 MeV, $f_{\pi}=$93 MeV and $g_{*s}$(present)$=$3.91, $f(\theta_1)$ is the anharmonic correction given in Fig. \ref{fig:combincorr} and $\gamma$ is the entropy increase ratio.

\begin{figure}[h]
\centering
\resizebox{0.85\columnwidth}{!}
{\includegraphics{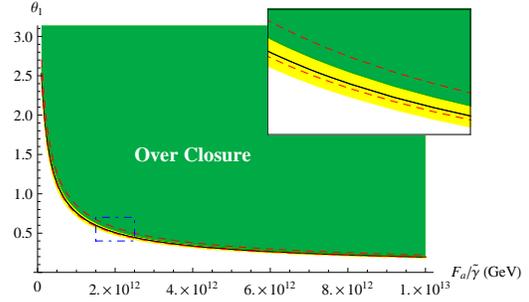}}
\caption{The bound from overclosure of the universe.  The yellow band shows the error bars of $\Lambda$ and two red dashed lines are the limits of the allowed current quark masses. Here, the entropy production ratio $\gamma$ is absorbed into the bracket of $F_a$: $\tilde\gamma=\gamma^{(n+4)/(n+6)}\simeq \gamma^{0.84}$.}
\label{fig:omega}
\end{figure}

The energy fraction of the axion is given by
\begin{eqnarray}
\Omega_a &\simeq& 0.379 \times \left(\frac{m_u m_d m_s}{3 \cdot 6 \cdot 103~ \rm MeV}\right)^{-0.092}\nonumber\\
&&\times\biggl(\frac{\theta_1^2F(\theta_1)}{\gamma}\biggr) \biggl(\frac{0.701}{h}\biggr)^2 \nonumber \\
&&\times \left(\frac{\Lambda_{\rm QCD}}{380\rm MeV}\right)^{-0.733}\left(\frac{F_a}{10^{12}{\rm GeV}}\right)^{1.184-0.010x},
\end{eqnarray}
where $x=(\Lambda_{\rm QCD}/380\rm MeV) -1$.
Of course, these expressions gives the bound given in Fig. \ref{fig:omega}.

\begin{theacknowledgments}
This work is supported in part by the Korea Research Foundation, Grant No. KRF-2005-084-C00001.
\end{theacknowledgments}

\bibliographystyle{aipproc}

\begin{thebibliography}{999}

\def\prp#1#2#3{Phys.\ Rep.\ {\bf #1}, #2 (#3)}
\def\rmp#1#2#3{Rev. Mod. Phys.\ {\bf #1}, #2 (#3)}
\def\anrnp#1#2#3{Annu. Rev. Nucl.
Part. Sci.\ {\bf #1}, #2 (#3)}
\def\npb#1#2#3{Nucl.\ Phys.\ {\bf B#1}, #2 (#3)}
\def\plb#1#2#3{Phys.\ Lett.\ {\bf B#1}, #2 (#3)}
\def\prd#1#2#3{Phys.\ Rev.\ {\bf D#1}, #2 (#3)}
\def\prl#1#2#3{Phys.\ Rev.\ Lett.\ {\bf #1}, #2 (#3)}
\def\jhep#1#2#3{JHEP\ {\bf #1}, #2 (#3)}
\def\jcap#1#2#3{JCAP\ {\bf #1}, #2 (#3)}
\def\zp#1#2#3{Z.\ Phys.\ {\bf #1} (#3) #2}
\def\epjc#1#2#3{Euro. Phys. J.\ {\bf C#1}, #2 (#3)}
\def\ijmp#1#2#3{Int.\ J.\ Mod.\ Phys.\ {\bf #1}, #2 (#3)}
\def\mpl#1#2#3{Mod.\ Phys.\ Lett.\ {\bf A#1}, #2 (#3)}
\def\apj#1#2#3{Astrophys.\ J.\ {\bf #1}, #2 (#3)}
\def\nat#1#2#3{Nature\ {\bf #1}, #2 (#3)}
\def\sjnp#1#2#3{Sov.\ J.\ Nucl.\ Phys.\ {\bf #1}, #2 (#3)}
\def\apj#1#2#3{Astrophys.\ J.\ {\bf #1}, #2 (#3)}
\def\ijmp#1#2#3{Int.\ J.\ Mod.\ Phys.\ {\bf #1}, #2 (#3)}
\def\mpl#1#2#3{Mod.\ Phys.\ Lett.\ {\bf A#1}, #2 (#3)}
\def\nat#1#2#3{Nature\ {\bf #1}, #2 (#3)}
\def\npb#1#2#3{Nucl.\ Phys.\ {\bf B#1}, #2 (#3)}
\def\apph#1#2#3{Astropart.\ Phys.\ {\bf B#1}, #2 (#3)}

\bibitem{Vafa:1984xg}  C.~Vafa and E.~Witten,  Phys.\ Rev.\ Lett.\  {\bf 53} (1984) 535.

\bibitem{Bae:2008ue} K.~J.~Bae, J.~H.~Huh and J.~E.~Kim, JCAP {\bf 0809} (2008) 005
  [arXiv:0806.0497 [hep-ph]].

\bibitem{Kim:2008hd} J.~E.~Kim and G.~Carosi,
  arXiv:0807.3125 [hep-ph].

\bibitem{PQ77} R. D. Peccei and H. R. Quinn, \prl{38}{1440}{1977}; S. Weinberg,\prl{40}{223}{1978}; F. Wilczek, \prl{40}{279}{1978}.

\bibitem{Komatsu:2008hk}
  E.~Komatsu {\it et al.}  [WMAP Collaboration],
   arXiv:0803.0547 [astro-ph].

\bibitem{Linde82} A. D. Linde  \plb{116}{335}{1982}.

\bibitem{Kim79} J. E. Kim, \prl{43}{103}{1979}; M. A. Shifman, V. I. Vainstein, V. I. Zakharov, \npb{166}{4933}{1980}; M. Dine, W. Fischler and M. Srednicki, \plb{104}{199}{1981}; A. P. Zhitnitskii, Sov. J. Nucl. Phys. {\bf 31}, 260 (1980).

\bibitem{Preskill83} J. Preskill, M. B. Wise and F. Wilczek, \plb{120}{127}{1983}; L. F. Abbott and P. Sikivie, \plb{120}{133}{1983}; M. Dine and W. Fischler, \plb{120}{137}{1983}.

\bibitem{Turner86} M. S. Turner, \prd{33}{889}{1986}.

\bibitem{DeGrand:1985uq}
  T.~A.~DeGrand, T.~W.~Kephart and T.~J.~Weiler,
   Phys.\ Rev.\  D {\bf 33}, 910 (1986).

\bibitem{Yaffe81} D. Gross, R. Pisarski and L. Yaffe, \rmp{53}{43}{1981}.

\bibitem{Hinchliffe02} I. Hinchliffe, \prd{66}{010001}{2002}.

\bibitem{Baluni79} V. Baluni, \prd{19}{2227}{1979}; W. A. Bardeen and S.-H. H. Tye,
    \plb{76}{580}{1978}.

\bibitem{Lyth91} D. H. Lyth, \prd{45}{3394}{1992}.

\bibitem{Landau} L. D. Landau and E. M. Lifshitz,  in {\em Mechanics} (Elsevier Butterworth-Heinemann, 1976), Sec. 49.

\bibitem{DeGrand:1984uq}
  T.~A.~DeGrand and K.~Kajantie,
   Phys.\ Lett.\  B {\bf 147}, 273 (1984).

\bibitem{Manohar06} A. V. Manohar and C. T. Sachrajda, in Particle Data Book [W.-M. Yao {\it et. al.}, J. of Physics {\bf G33}, 1 (2006)].

\bibitem{Linde05} A.~D.~Linde, arXiv:hep-th/0503203.

\end{thebibliography}

\end{document}